\documentclass[12pt]{article}
\usepackage{amsmath,amsfonts,amssymb}  

\makeatletter
\def\numberbysection{\@addtoreset{equation}{section}
    \def\theequation{\thesection.\arabic{equation}}}
\makeatother

\numberbysection

\newcommand{\be}{\begin{eqnarray}}
\newcommand{\ee}{\end{eqnarray}}
\newcommand{\non}{\nonumber}
\newcommand{\id}{\mathbb{I}}

\newcommand{\diag}{\mathop{\rm diag}\nolimits}

\begin{document}

\begin{titlepage}
\strut\hfill UMTG--257
\vspace{.5in}
\begin{center}

\LARGE $q$-deformed $su(2|2)$ boundary $S$-matrices\\
\LARGE via the ZF algebra\\[1.0in]
\large Rajan Murgan and Rafael I. Nepomechie\\[0.8in]
\large Physics Department, P.O. Box 248046, University of Miami\\[0.2in]  
\large Coral Gables, FL 33124 USA\\
      
\end{center}

\vspace{.5in}

\begin{abstract}
Beisert and Koroteev have recently found a bulk $S$-matrix
corresponding to a $q$-deformation of the centrally-extended $su(2|2)$
algebra of AdS/CFT. We formulate the associated Zamolodchikov-Faddeev
algebra, using which we derive factorizable boundary $S$-matrices that
generalize those of Hofman and Maldacena.
\end{abstract}

\end{titlepage}

\setcounter{footnote}{0}

\section{Introduction}\label{sec:intro}

In investigations of integrability in AdS/CFT (for reviews, see for
example \cite{reviews}), a centrally-extended $su(2|2)$ algebra (more
precisely, two such copies) has emerged as a key symmetry: it is the
residual symmetry algebra of both planar ${\cal N}=4$ super Yang-Mills
theory \cite{Be} and the gauge-fixed $AdS_{5}\times S^{5}$ superstring
sigma model \cite{AFPZ}.  This symmetry leads directly to a bulk
$S$-matrix \cite{Be} for the fundamental excitations \cite{BMN} which
obeys a twisted (dynamical) Yang-Baxter equation.  By introducing
Zamolodchikov-Faddeev (ZF) operators \cite{ZZ, Fa} that have suitable
commutation relations with the symmetry generators, a related bulk
$S$-matrix can be derived \cite{AFZ} which obeys the standard
Yang-Baxter equation.  These $S$-matrices have been used to prove
\cite{Be, MM, dL} a previously-conjectured set of asymptotic Bethe
equations \cite{BS} for the spectrum of the gauge/string theory.

Some of these results have been generalized to the case where there is
a boundary.  Hofman and Maldacena \cite{HM} have considered open
strings attached to maximal giant gravitons \cite{giants} in
$AdS_{5}\times S^{5}$.  (See also \cite{BV, Ag, MV} and references
therein.)  Based on the residual symmetries, they have derived
corresponding boundary $S$-matrices.  By extending the ZF algebra 
\cite{AFZ} to the boundary case, related boundary $S$-matrices
which obey the standard boundary Yang-Baxter equation \cite{Ch, GZ}
have been derived in \cite{AN}.

A $q$-deformation of this centrally-extended $su(2|2)$ algebra has
recently been considered by Beisert and Koroteev \cite{BK}. They 
derived a corresponding bulk $S$-matrix, which they related to a 
deformation \cite{AB} of the one-dimensional Hubbard model \cite{EFGKK}.

In this note, we formulate the ZF algebra associated with this
deformed symmetry algebra, using which we derive corresponding
factorizable boundary $S$-matrices.  The ZF formalism is particularly
convenient for performing explicit calculations, as the coproduct and
braiding relations are encoded in the commutation relations of the ZF
operators with the symmetry generators.  Using these deformed bulk and
boundary $S$-matrices as inputs into Sklyanin's generalization of the
Quantum Inverse Scattering Method for systems with boundaries
\cite{Sk}, it should be possible to construct and solve open versions
of the deformed Hubbard model.  However, even for the undeformed case,
this problem remains a challenge.

This paper is organized as follows.  In Section \ref{sec:algebra} we
recall the definition of the $q$-deformed centrally-extended $su(2|2)$
algebra \cite{BK}.  In Section \ref{sec:bulk}, we introduce the bulk ZF
algebra, and present the commutation relations of the ZF operators
with the symmetry generators.  As a check on these relations, we use
them to recover the Beisert-Koroteev $S$-matrix.  We address boundary
scattering in Section \ref{sec:boundary}.  We begin by determining
how $x^{\pm}$ transforms under the reflection $p \mapsto -p$. We then 
extend the ZF algebra by introducing suitable boundary operators, and 
proceed to construct $q$-deformations of the $Y=0$ and $Z=0$ giant
graviton brane boundary $S$-matrices of Hofman and Maldacena.

\section{The $q$-deformed algebra}\label{sec:algebra}

We briefly review here the $q$-deformed centrally-extended $su(2|2)$
algebra.  Following \cite{BK}, we work in the Chevalley basis, with
three Cartan generators $h_{j}$, three simple positive roots $E_{j}$
and three simple negative roots $F_{j}$, $j=1, 2, 3$.  The generators
$E_{2}\,, F_{2}$ are fermionic, while the remaining ones are bosonic.
The commutators with the Cartan generators are given by
\be
\left[ h_{j}\,, h_{k} \right] = 0\,, \qquad
\left[ h_{j}\,, E_{k} \right] = A_{jk}\, E_{k}\,, \qquad
\left[ h_{j}\,, F_{k} \right] = -A_{jk}\, F_{k}\,, 
\label{Cartan}
\ee
where $A_{jk}$ is the symmetric Cartan matrix
\be
A_{jk} = \left(\begin{array}{rrr}
2 &-1 &0 \\
-1 &0  &1 \\
0 &1 &-2
\end{array} \right) \,.
\ee
The commutators of $E_{j}$ with $F_{j}$ are given by 
\be
\left[ E_{1}\,, F_{1} \right] = \left[ h_{1} \right]_{q}\,, \qquad
\left\{ E_{2}\,, F_{2} \right\} = -\left[ h_{2} \right]_{q}\,, \qquad
\left[ E_{3}\,, F_{3} \right] = -\left[ h_{3} \right]_{q}\,, 
\label{EF1}
\ee
where
\be
\left[x \right]_{q} = \frac{q^{x}-q^{-x}}{q-q^{-1}} \,,
\ee
and $q$ is the deformation parameter. The remaining mixed commutators 
vanish,
\be
\left[ E_{j}\,, F_{k} \right] = 0\,, \qquad j \ne k \,.
\label{EF2}
\ee
The Serre relations are given by
\be
& & \left[ E_{1}\,, E_{3} \right] = \left[ F_{1}\,, F_{3} \right] = 
E_{2} E_{2} = F_{2} F_{2} = 0 \,, \label{Serre} \\
& & E_{1} E_{1} E_{2} - (q+q^{-1}) E_{1} E_{2} E_{1} + E_{2} E_{1} E_{1} =
E_{3} E_{3} E_{2} - (q+q^{-1}) E_{3} E_{2} E_{3} + E_{2} E_{3} E_{3} 
=  0 \,, \non \\
& & F_{1} F_{1} F_{2} - (q+q^{-1}) F_{1} F_{2} F_{1} + F_{2} F_{1} F_{1} =
F_{3} F_{3} F_{2} - (q+q^{-1}) F_{3} F_{2} F_{3} + F_{2} F_{3} F_{3} 
=  0 \,. \non 
\ee 
The algebra has three central elements given by
\be
C &=& -\frac{1}{2}h_{1} -h_{2} -\frac{1}{2}h_{3} \,, \non \\
P &=& E_{1} E_{2} E_{3} E_{2} + E_{2} E_{3} E_{2} E_{1} 
+ E_{3} E_{2} E_{1} E_{2} +  E_{2} E_{1} E_{2} E_{3} 
- (q+q^{-1}) E_{2} E_{1} E_{3} E_{2}  \,, \non \\
K &=& F_{1} F_{2} F_{3} F_{2} + F_{2} F_{3} F_{2} F_{1} 
+ F_{3} F_{2} F_{1} F_{2} +  F_{2} F_{1} F_{2} F_{3} 
- (q+q^{-1}) F_{2} F_{1} F_{3} F_{2}  \,.
\label{centralcharges}
\ee

\section{Bulk scattering}\label{sec:bulk}

We introduce here the bulk ZF algebra, and present the commutation
relations of the ZF operators with the symmetry generators.  As a
check on these relations, we then verify that the Beisert-Koroteev
$S$-matrix can be recovered by demanding that the symmetry generators
commute with two-particle scattering.

\subsection{Bulk ZF algebra}\label{subsec:bulkZF} 

Following \cite{AFZ, AN}, we
denote the ZF operators by $A_{i}^{\dagger}(p)$, $i=1\,, 2\,, 3\,, 4$.
These operators create asymptotic particle states of momentum $p$ when
acting on the vacuum state $|0\rangle$, corresponding to 
$|\phi^{1}\rangle$, $|\phi^{2}\rangle$, $|\psi^{1}\rangle$, 
$|\psi^{2}\rangle$ in \cite{BK}, respectively.
Hence, the first two operators are bosonic, 
while the last two operators are fermionic.
The bulk $S$-matrix elements
$S_{i\, j}^{i' j'}(p_{1}, p_{2})$ are defined by the relation
\be
A_{i}^{\dagger}(p_{1})\, A_{j}^{\dagger}(p_{2}) = 
S_{i\, j}^{i' j'}(p_{1}, p_{2})\, 
A_{j'}^{\dagger}(p_{2})\, A_{i'}^{\dagger}(p_{1}) \,,
\label{bulkS1}
\ee
where summation over repeated indices is always understood.  It is
convenient to arrange these matrix elements into a $16 \times 16$
matrix $S$ as follows,
\be
S = S_{i\, j}^{i' j'} e_{i\, i'}\otimes e_{j\, j'}\,,
\label{bulkS2}
\ee
where $e_{i j}$ is the usual elementary $4 \times 4$ matrix whose 
$(i, j)$ matrix element is 1, and all others are zero.
Associativity of the  ZF algebra implies \cite{ZZ} the Yang-Baxter equation,
\be
S_{12}(p_{1}, p_{2})\, S_{13}(p_{1}, p_{3})\, S_{23}(p_{2}, p_{3})\ =
S_{23}(p_{2}, p_{3})\, S_{13}(p_{1}, p_{3})\, S_{12}(p_{1}, p_{2}) \,.
\label{YBE}
\ee
We use the standard convention $S_{12} = S \otimes \id$, $S_{23} 
= \id \otimes S$, and $S_{13} = {\cal P}_{12}\, S_{23}\, {\cal P}_{12}$,
where ${\cal P}_{12} = {\cal P} \otimes \id$, ${\cal P} = e_{i\, j} \otimes
e_{j\, i}$ is the permutation matrix, and $\id$ is the four-dimensional
identity matrix. 

The one-particle states $A_{i}^{\dagger}(p) |0\rangle$ must form a
fundamental representation of the symmetry algebra (see Eq.  (2.55) in
\cite{BK}); and similarly, multi-particle states must form higher
(reducible) representations.  From these requirements, and the fact
that the symmetry generators annihilate the vacuum state, we can
abstract the action of the symmetry generators on the ZF operators.

The nontrivial commutators of the Cartan generators 
with the ZF operators are given by
\be
h_{1}\, A_{1}^{\dagger}(p) &=& -A_{1}^{\dagger}(p) +  
A_{1}^{\dagger}(p)\, h_{1}\,, \qquad
h_{1}\, A_{2}^{\dagger}(p) = A_{2}^{\dagger}(p) +  
A_{2}^{\dagger}(p)\, h_{1}\,,  \non \\
h_{3}\, A_{3}^{\dagger}(p) &=& -A_{3}^{\dagger}(p) +  
A_{3}^{\dagger}(p)\, h_{3}\,,  \qquad
h_{3}\, A_{4}^{\dagger}(p) = A_{4}^{\dagger}(p) +  
A_{4}^{\dagger}(p)\, h_{3}\,,   \label{CartanZF} \\
h_{2}\, A_{1}^{\dagger}(p) &=& -\big(C-\frac{1}{2}\big) A_{1}^{\dagger}(p) +  
A_{1}^{\dagger}(p)\, h_{2}\,, \qquad
h_{2}\, A_{2}^{\dagger}(p) = -\big(C+\frac{1}{2}\big) A_{2}^{\dagger}(p) +  
A_{2}^{\dagger}(p)\, h_{2}\,,  \non \\
h_{2}\, A_{3}^{\dagger}(p) &=& -\big(C-\frac{1}{2}\big) A_{3}^{\dagger}(p) +  
A_{3}^{\dagger}(p)\, h_{2}\,, \qquad
h_{2}\, A_{4}^{\dagger}(p) = -\big(C+\frac{1}{2}\big) A_{4}^{\dagger}(p) +  
A_{4}^{\dagger}(p)\, h_{2}\,,  \non 
\ee
where $C=C(p)$ denotes the value of the corresponding central 
charge $C$  (\ref{centralcharges}). The remaining such commutators 
are trivial, $h_{j}\, A_{k}^{\dagger}(p) =   A_{k}^{\dagger}(p)\, 
h_{j}$.

The nontrivial commutators of the bosonic simple roots with the ZF
operators are given by
\be
E_{1}\, A_{1}^{\dagger}(p) &=& q^{1/2}\, A_{2}^{\dagger}(p)\, 
q^{-h_{1}/2} +  q^{-1/2}\, A_{1}^{\dagger}(p)\, E_{1}\,, \qquad
E_{1}\, A_{2}^{\dagger}(p) = q^{1/2}\, A_{2}^{\dagger}(p)\, E_{1}\,, 
\non \\
E_{3}\, A_{4}^{\dagger}(p) &=& q^{-1/2}\, A_{3}^{\dagger}(p)\, 
q^{-h_{3}/2} +  q^{1/2}\, A_{4}^{\dagger}(p)\, E_{3}\,, \qquad
E_{3}\, A_{3}^{\dagger}(p) = q^{-1/2}\, A_{3}^{\dagger}(p)\, E_{3}\,, 
\non \\
F_{1}\, A_{2}^{\dagger}(p) &=& q^{-1/2}\, A_{1}^{\dagger}(p)\, 
q^{-h_{1}/2} +  q^{1/2}\, A_{2}^{\dagger}(p)\, F_{1}\,, \qquad
F_{1}\, A_{1}^{\dagger}(p) = q^{-1/2}\, A_{1}^{\dagger}(p)\, F_{1}\,, 
\non \\
F_{3}\, A_{3}^{\dagger}(p) &=& q^{1/2}\, A_{4}^{\dagger}(p)\, 
q^{-h_{3}/2} +  q^{-1/2}\, A_{3}^{\dagger}(p)\, F_{3}\,, \qquad
F_{3}\, A_{4}^{\dagger}(p) = q^{1/2}\, A_{4}^{\dagger}(p)\, F_{3}\,.
\label{bosonicsimpleZF}
\ee
The remaining such commutators are trivial, namely,
\be
E_{1}\, A_{\alpha}^{\dagger}(p) &=& A_{\alpha}^{\dagger}(p)\, E_{1}\,, \qquad
F_{1}\, A_{\alpha}^{\dagger}(p) = A_{\alpha}^{\dagger}(p)\, F_{1}\,, \qquad
\alpha = 3, 4 \,, \non \\
E_{3}\, A_{a}^{\dagger}(p) &=& A_{a}^{\dagger}(p)\, E_{3}\,, \qquad
F_{3}\, A_{a}^{\dagger}(p) = A_{a}^{\dagger}(p)\, F_{3}\,, \qquad
a = 1, 2 \,.
\ee

Finally, the commutators of the fermionic generators with the ZF
operators are given by
\be
E_{2}\, A_{2}^{\dagger}(p) &=& e^{-i p/2}\left[
a(p)\, A_{4}^{\dagger}(p)\, 
q^{-h_{2}/2} +  q^{-(C+\frac{1}{2})/2}\, A_{2}^{\dagger}(p)\, E_{2} 
\right]\,, \non \\
E_{2}\, A_{3}^{\dagger}(p) &=& e^{-i p/2}\left[
b(p)\, A_{1}^{\dagger}(p)\, 
q^{-h_{2}/2} -  q^{-(C-\frac{1}{2})/2}\, A_{3}^{\dagger}(p)\, E_{2} 
\right]\,, \non \\
F_{2}\, A_{1}^{\dagger}(p) &=& e^{i p/2}\left[
c(p)\, A_{3}^{\dagger}(p)\, 
q^{-h_{2}/2} +  q^{-(C-\frac{1}{2})/2}\, A_{1}^{\dagger}(p)\, F_{2} 
\right]\,, \non \\
F_{2}\, A_{4}^{\dagger}(p) &=& e^{i p/2}\left[
d(p)\, A_{2}^{\dagger}(p)\, 
q^{-h_{2}/2} -  q^{-(C+\frac{1}{2})/2}\, A_{4}^{\dagger}(p)\, F_{2} 
\right]\,, 
\label{fermionicZF1}
\ee
and
\be
E_{2}\, A_{1}^{\dagger}(p) &=& e^{-i p/2} q^{-(C-\frac{1}{2})/2}\, 
A_{1}^{\dagger}(p)\, E_{2} \,,  \qquad 
E_{2}\, A_{4}^{\dagger}(p) = -e^{-i p/2} q^{-(C+\frac{1}{2})/2}\, 
A_{4}^{\dagger}(p)\, E_{2} \,, \non \\
F_{2}\, A_{2}^{\dagger}(p) &=& e^{i p/2} q^{-(C+\frac{1}{2})/2}\, 
A_{2}^{\dagger}(p)\, F_{2} \,,  \qquad 
F_{2}\, A_{3}^{\dagger}(p) = -e^{i p/2} q^{-(C-\frac{1}{2})/2}\, 
A_{3}^{\dagger}(p)\, F_{2} \,.
\label{fermionicZF2}
\ee
The one-particle states form a representation of the algebra with 
$P=e^{-i p}a b$, $K=e^{i p} c d$, provided the functions $a, b, c, d$ 
obey the constraints \cite{BK}
\be
a d =\big[C + \frac{1}{2}\big]_{q}\,, 
\quad b c = \big[C - \frac{1}{2}\big]_{q}\,,
\label{constraints}
\ee
which imply
\be
(a d - q b c)(a d - q^{-1} b c) = 1 \,.
\ee

We have verified that the above commutation relations are consistent with 
the symmetry algebra (\ref{Cartan}) - (\ref{Serre}).  
Notice the appearance of the Cartan generators $h_{j}$ in the commutation 
relations (\ref{bosonicsimpleZF}),  (\ref{fermionicZF1}), which is 
necessary to implement the nontrivial coproduct. (See, for example, 
\cite{BL}.)

A further constraint on the functions $a, b, c, d$ comes from the
requirement \cite{BK} that the central charges $P$ and $K$ (\ref{centralcharges}) 
commute with two-particle scattering.  Indeed, acting with $P$ on 
both sides of \footnote{The $S$-matrix elements $S_{1\, 1}^{i\, j}$ with $i, j \ne 
1$ vanish, as can be seen from (\ref{bulkS3}) below.}
\be
A_{1}^{\dagger}(p_{1})\, A_{1}^{\dagger}(p_{2})\, |0\rangle =
S_{1\, 1}^{1\, 1}(p_{1}, p_{2})\, A_{1}^{\dagger}(p_{2})\, 
A_{1}^{\dagger}(p_{1})\, |0\rangle \,,
\label{11state}
\ee
we obtain
\be
e^{-i p_{1}} q^{C_{2}} a_{1} b_{1} + e^{-i (p_{1}+p_{2})} q^{-C_{1}} 
a_{2} b_{2} = e^{-i p_{2}} q^{C_{1}} a_{2} b_{2} + e^{-i 
(p_{1}+p_{2})} q^{-C_{2}} a_{1} b_{1} \,,
\ee
which implies 
\be
\frac{a_{1} b_{1}}{q^{C_{1}} e^{i p_{1}} - q^{-C_{1}}} =
\frac{a_{2} b_{2}}{q^{C_{2}} e^{i p_{2}} - q^{-C_{2}}} = 
constant \,.
\label{Pconstraint}
\ee
Similarly, acting with $K$ on (\ref{11state}), we obtain 
\be
\frac{c_{1} d_{1}}{q^{C_{1}} e^{-i p_{1}} - q^{-C_{1}}} =
\frac{c_{2} d_{2}}{q^{C_{2}} e^{-i p_{2}} - q^{-C_{2}}} = 
constant \,.
\label{Kconstraint}
\ee

The constraints (\ref{Pconstraint}) and (\ref{Kconstraint}) are 
satisfied if we set \footnote{Our expressions for $a$ and $d$ differ 
from those in \cite{BK} by factors of $q^{\mp C}$. Also, Beisert and Koroteev 
do not introduce a momentum variable $p$; instead, they work with $U 
= e^{i p/2}$.}
\be
a &=& \sqrt{g}\, \gamma\, q^{-C} \,, \non \\
b &=& \sqrt{g}\, \frac{\alpha}{\gamma}\, \frac{1}{x^{-}}
\left(x^{-} - q^{2C-1} x^{+}\right)  \,, \non \\
c &=& i\sqrt{g}\, \frac{\gamma}{\alpha}\, \frac{q^{-C+\frac{1}{2}}}{x^{+}}\,, \non \\
d &=& i\sqrt{g}\,  \frac{q^{-\frac{1}{2}}}{\gamma}
\left( q^{2C+1} x^{-} -  x^{+}\right)\,, 
\ee 
with 
\be
e^{i p} = \frac{x^{+}}{q x^{-}} \,.
\label{momentum}
\ee
For simplicity, we henceforth set $\alpha = 1$; but (as in \cite{BK}) 
we leave $\gamma$ unspecified.

The constraints (\ref{constraints}) then imply \cite{BK}
\be
q^{2C} = \frac{1}{q} \left(\frac{1 - i g (q - q^{-1}) x^{+}}
{1 - i g (q - q^{-1}) x^{-}} \right) = 
q  \left(\frac{1 +  i g (q - q^{-1})/x^{+}}
{1 + i g (q - q^{-1})/x^{-}} \right) \,.
\label{qC}
\ee
These relations in turn imply the quadratic constraint \cite{BK}
\be
\frac{x^{+}}{q} +\frac{q}{x^{+}} - q x^{-} - \frac{1}{q x^{-}} 
+ i g (q - q^{-1})\left(\frac{x^{+}}{q x^{-}} -\frac{q 
x^{-}}{x^{+}}\right) = \frac{i}{g} \,.
\label{quadratic}
\ee 

\subsection{Bulk $S$-matrix}\label{subsec:bulkS}

As usual, we can determine the two-particle $S$-matrix (up to a phase)
by demanding that the symmetry generators commute with two-particle
scattering.  That is, starting from
$J\, A_{i}^{\dagger}(p_{1})\, A_{j}^{\dagger}(p_{2})
|0\rangle$ where $J$ is a symmetry generator, and assuming
that $J$ annihilates the vacuum state, we
arrive at linear combinations of $A_{j'}^{\dagger}(p_{2})\,
A_{i'}^{\dagger}(p_{1}) |0\rangle$ in two different ways, by applying the
ZF relation (\ref{bulkS1}) and the symmetry relations 
(\ref{CartanZF}) - (\ref{fermionicZF2})
in different orders.  The consistency condition is
a system of linear equations for the $S$-matrix
elements.  The result for the nonzero matrix elements is
\be
S_{a\, a}^{a\, a} &=& \mathcal{ A}\,, \qquad \qquad \ \
S_{\alpha\, \alpha}^{\alpha\, \alpha} = \mathcal{ D}\,, \non \\
S_{a\, b}^{a\, b} &=& 
\frac{\mathcal{ A}-\mathcal{ B}}{q+q^{-1}}\,, \qquad 
S_{a\, b}^{b\, a} = \frac{q^{-\epsilon_{a b}}\mathcal{ A}
+q^{\epsilon_{a b}}\mathcal{ B}}{q+q^{-1}} \,, \non \\
S_{\alpha\, \beta}^{\alpha\, \beta} &=& 
\frac{\mathcal{ D}-\mathcal{ E}}{q+q^{-1}}\,, \quad 
S_{\alpha\, \beta}^{\beta\, \alpha} = 
\frac{q^{-\epsilon_{\alpha \beta}}\mathcal{ D}
+q^{\epsilon_{\alpha \beta}}\mathcal{ E}}{q+q^{-1}} \,, \non \\
S_{a\, b}^{\alpha\, \beta} &=& 
q^{(\epsilon_{a b}-\epsilon_{\alpha \beta})/2}
\epsilon_{a b}\epsilon^{\alpha \beta}\, \frac{\mathcal{ C}}{q+q^{-1}} \,, \quad
S_{\alpha\, \beta}^{a\, b} = 
q^{(\epsilon_{\alpha \beta}-\epsilon_{a b})/2}
\epsilon^{a b}\epsilon_{\alpha \beta}\, \frac{\mathcal{ F}}{q+q^{-1}} \,, \non \\
S_{a\, \alpha}^{a\, \alpha} &=& \mathcal{ L}\,, \quad 
S_{a\, \alpha}^{\alpha\, a} = \mathcal{ K} \,, \quad 
S_{\alpha\, a}^{a\, \alpha} = \mathcal{ H}\,, \quad 
S_{\alpha\, a}^{\alpha\, a} = \mathcal{ G} \,,  
\label{bulkS3}
\ee
where $a\,, b \in \{1\,, 2\}$ with $a \ne b$;  
$\alpha\,, \beta \in \{3\,, 4\}$ with $\alpha \ne \beta$; and
\be
\mathcal{ A} &=& A^{BK}_{21} = 
q^{C_{2}-C_{1}}e^{i(p_{2}-p_{1})/2}
\frac{x^{+}_{1}-x^{-}_{2}}{x^{-}_{1}-x^{+}_{2}}\,, \non \\
\mathcal{ B} &=& B^{BK}_{21} = 
q^{C_{2}-C_{1}}e^{i(p_{2}-p_{1})/2}\frac{x^{+}_{1}-x^{-}_{2}}{x^{-}_{1}-x^{+}_{2}}
\left(1-(q+q^{-1})q^{-1}\frac{x^{+}_{1}-x^{+}_{2}}{x^{+}_{1}-x^{-}_{2}}
\frac{x^{-}_{1}-s(x^{+}_{2})}{x^{-}_{1}-s(x^{-}_{2})}\right)
\,, \non \\
\mathcal{ C} &=& q^{-(C_{1}+C_{2}-1)/2}C^{BK}_{21} = (q+q^{-1})
i g q^{(C_{2}-5C_{1}-2)/2}e^{i (p_{2}-2p_{1})/2} \gamma_{1}\gamma_{2}
\frac{i g^{-1}x^{+}_{1}-(q-q^{-1})}{x^{-}_{1}-s(x^{-}_{2})}
\frac{s(x^{+}_{1})-s(x^{+}_{2})}{x^{-}_{1}-x^{+}_{2}}
\,, \non
\ee
\be 
\mathcal{ D} &=& -1 \,, \non \\
\mathcal{ E} &=& E^{BK}_{21} = -\left(1-(q+q^{-1})q^{-2C_{1}-1}e^{-i p_{1}}
\frac{x^{+}_{1}-x^{+}_{2}}{x^{-}_{1}-x^{+}_{2}}
\frac{x^{+}_{1}-s(x^{-}_{2})}{x^{-}_{1}-s(x^{-}_{2})}\right)
\,, \non \\
\mathcal{ F} &=& q^{(C_{1}+C_{2}-1)/2}F^{BK}_{21} = -(q+q^{-1})
i g q^{(5C_{2}-C_{1}-2)/2}e^{i (2p_{2}-p_{1})/2}
\frac{i g^{-1}x^{+}_{1}-(q-q^{-1})}{x^{-}_{1}-s(x^{-}_{2})}
\frac{s(x^{+}_{1})-s(x^{+}_{2})}{x^{-}_{1}-x^{+}_{2}}\non \\
& & \cdot \frac{1}{1-g^{2}(q+q^{-1})^{2}} \frac{1}{\gamma_{1}\gamma_{2}}
(x^{+}_{1}-x^{-}_{1})(x^{+}_{2}-x^{-}_{2})
\,, \non \\
\mathcal{ G} &=& G^{BK}_{21} = q^{-C_{1}-1/2} e^{-i p_{1}/2} 
\frac{x^{+}_{1}-x^{+}_{2}}{x^{-}_{1}-x^{+}_{2}}
\,, \non \\
\mathcal{ H} &=& q^{(C_{1}-C_{2})/2}H^{BK}_{21} = q^{(C_{1}-C_{2})/2}
\frac{\gamma_{2}}{\gamma_{1}}
\frac{x^{+}_{1}-x^{-}_{1}}{x^{-}_{1}-x^{+}_{2}}
\,, \non \\
\mathcal{ K} &=& q^{-(C_{1}-C_{2})/2}K^{BK}_{21} = q^{3(C_{2}-C_{1})/2}
e^{i(p_{2}-p_{1})/2}\frac{\gamma_{1}}{\gamma_{2}}
\frac{x^{+}_{2}-x^{-}_{2}}{x^{-}_{1}-x^{+}_{2}}
\,, \non \\
\mathcal{ L} &=& L^{BK}_{21}= q^{C_{2}+1/2} e^{i p_{2}/2}
\frac{x^{-}_{1}-x^{-}_{2}}{x^{-}_{1}-x^{+}_{2}}
\,, \label{bulkS4}
\ee 
where $A^{BK}_{21}\,, B^{BK}_{21}\,, \ldots$ denote the amplitudes
$A_{12}\,, B_{12}\,, \ldots$ in Table 2 of \cite{BK}, respectively,
with labels 1 and 2 interchanged.  As already mentioned, we have set
the parameter $\alpha$, as well as the overall scalar
factor (denoted by $R^{0}$ in \cite{BK}), equal to one.  The function
$s(x)$ is the ``antipode map'' defined by \cite{BK}
\be
s(x) = \frac{1 - ig (q-q^{-1}) x}{x + ig (q-q^{-1})} \,,
\label{sfunc}
\ee
which has the limit $s(x) \rightarrow 1/x$ for $q \rightarrow 1$.  Our
amplitudes $\mathcal{ C}$, $\mathcal{ F}$, $\mathcal{ H}$ and
$\mathcal{ K}$ evidently have extra factors involving powers of $q$
with respect to the amplitudes of Beisert and Koroteev.  However, we
have verified with Mathematica that the $S$-matrix satisfies the
Yang-Baxter equation (\ref{YBE}) even without those extra factors.
Hence, these factors can presumably be removed by a suitable gauge
transformation.

\section{Boundary scattering}\label{sec:boundary}

A prerequisite to studying boundary scattering is to understand how 
$x^{\pm}$ transforms under the reflection $p \mapsto -p$. We claim that
\be
x^{+}(-p) =-\frac{1}{s(x^{-}(p))}\,, \qquad
x^{-}(-p) =-\frac{1}{s(x^{+}(p))}\,,
\label{reflection}
\ee
where $s(x)$ is given by (\ref{sfunc}).  Indeed, (\ref{reflection})
has the correct $q \rightarrow 1$ limit, namely, $x^{\pm}(-p) =
-x^{\mp}(p)$ \cite{HM}.  Moreover, the momentum relation
(\ref{momentum}) is preserved by this transformation
\be
e^{-i p} = \frac{x^{+}(-p)}{q x^{-}(-p)} =  \frac{s(x^{+}(p))}{q 
s(x^{-}(p))} = \frac{q x^{-}(p)}{x^{+}(p)} \,,
\ee
where the final equality is an identity which can be found in Appendix
A of \cite{BK}.  Also, the transformation (\ref{reflection}) preserves
the energy,
\be
C(-p) = C(p) \,.
\label{energypreserved}
\ee
This can easily be seen as follows, starting from (\ref{qC}),
\be
q^{2C(-p)} &=&
q  \left(\frac{1 +  i g (q - q^{-1})/x^{+}(-p)}
{1 + i g (q - q^{-1})/x^{-}(-p)} \right) 
= q  \left(\frac{1 - i g (q - q^{-1})\, s(x^{-}(p))}
{1 - i g (q - q^{-1})\, s(x^{+}(p))} \right) \non \\
&=& q  \left(\frac{1 +  i g (q - q^{-1})/x^{+}(p)}
{1 + i g (q - q^{-1})/x^{-}(p)} \right) = q^{2C(p)}
\,,
\ee
where the first equality on the second line follows from the identity
\be
\frac{1 - i g (q - q^{-1})\, s(x)}
{1 - i g (q - q^{-1})\, s(y)}  =
\frac{1 +  i g (q - q^{-1})/y}
{1 + i g (q - q^{-1})/x} 
\ee
which holds for arbitrary values of $x$ and $y$. Furthermore, the 
transformation (\ref{reflection}) preserves the
quadratic relation (\ref{quadratic}), since
\be
-\frac{1}{q s(x^{-})} -q s(x^{-}) + \frac{q}{s(x^{+})} + \frac{s(x^{+})}{q} + 
i g (q - q^{-1})\left[\frac{s(x^{+})}{q s(x^{-})} - 
\frac{q s(x^{-})}{s(x^{+})}\right] = \frac{i}{g} \,,
\ee 
which can be seen most readily from the relation (\ref{qC}) and the 
fact (\ref{energypreserved}). Finally, we verify that the 
transformation (\ref{reflection}) squares to the identity, by virtue 
of the identity
\be
-s\left(-\frac{1}{s(x)}\right) = \frac{1}{x} \,,
\ee
which holds for arbitrary values of $x$.

Having determined how $x^{\pm}$ transforms under reflection, we turn
now to the problem of computing boundary $S$-matrices.  Following the
approach in \cite{AN}, we shall extend the ZF algebra (\ref{bulkS1})
by introducing suitable boundary operators which create the
boundary-theory vacuum state $|0\rangle_{B}$ from $|0\rangle$
\cite{GZ}.  We shall then proceed, using the commutation relations of
the ZF operators with the symmetry generators found in the previous
Section, to construct $q$-deformations of the $Y=0$ and $Z=0$ giant
graviton brane boundary $S$-matrices of Hofman and Maldacena
\cite{HM}.

\subsection{$Y=0$ giant graviton brane}\label{subsec:Y}

Since there is no boundary degree of freedom
for the $Y=0$ giant graviton brane, the corresponding boundary
operator is a scalar, $B$. The boundary $S$-matrix is defined by 
\footnote{We restrict our attention to the {\it right} boundary $S$-matrix,
since the left boundary $S$-matrix can be obtained by $p \mapsto -p$
\cite{HM, AN}.}
\be
A^{\dagger}_{i}(p)\, B = R_{i}^{\, i'}(p)\, A^{\dagger}_{i'}(-p)\,  
B \,.
\label{boundarySRY}
\ee
We arrange the $S$-matrix elements in the usual way into a matrix 
$R = R_{i}^{\, i'}\, e_{i\, i'}$.
Starting from $A^{\dagger}_{i}(p_{1})\, A^{\dagger}_{j}(p_{2})\, B$, 
one can arrive at linear combinations of 
$A^{\dagger}_{i'''}(-p_{1})\, A^{\dagger}_{j'''}(-p_{2})\, B$
by applying each of the relations (\ref{bulkS1}) and 
(\ref{boundarySRY}) two times, in two different ways. The consistency 
condition is the BYBE \cite{Ch, GZ}
\be
S_{12}(p_{1}, p_{2})\, R_{1}(p_{1})\, S_{21}(p_{2}, -p_{1})\, 
R_{2}(p_{2}) = 
R_{2}(p_{2})\, S_{12}(p_{1}, -p_{2})\, R_{1}(p_{1})\, 
S_{21}(-p_{2}, -p_{1}) \,.
\label{BYBERY}
\ee 

Let us assume that the vacuum state $B |0\rangle$ breaks $E_{1}\,,
F_{1}$, but preserves $E_{3}\,, F_{3}$.  It follows from
(\ref{bosonicsimpleZF}) that the boundary $S$-matrix is diagonal, with
matrix elements
\be 
R_{1}^{\, 1} = r_{1}\,, \quad R_{2}^{\, 2} = r_{2}\,, 
\quad R_{3}^{\, 3} =  R_{4}^{\, 4} = r \,. 
\ee
Using first  (\ref{fermionicZF1}) and then (\ref{boundarySRY}), we find
\be
E_{2}\,  A^{\dagger}_{2}(p)\, B |0\rangle
= e^{-i p/2}  a(p)  A^{\dagger}_{4}(p)  B |0\rangle = e^{-i p/2} a(p) r  
A^{\dagger}_{4}(-p) B |0\rangle\,,
\label{result1}
\ee
where we have passed to the second equality using also the assumption
that $E_{2}$ annihilates the vacuum state.  Reversing the order, i.e.,
using first (\ref{boundarySRY}) and then (\ref{fermionicZF1}), we
obtain
\be
E_{2}\,  A^{\dagger}_{2}(p)\, B |0\rangle
= r_{2} E_{2}\,  A^{\dagger}_{2}(-p)\, B |0\rangle
= r_{2} e^{i p/2}  a(-p)  A^{\dagger}_{4}(-p)\, B |0\rangle \,.
\label{result2}
\ee
Consistency of the results (\ref{result1}) and (\ref{result2}) requires
\be
\frac{r_{2}}{r} = e^{-i p} \frac{a(p)}{a(-p)} 
= e^{-i p} \frac{d(-p)}{d(p)} 
= e^{-i p}
\frac{\gamma(p)}{\gamma(-p)}\,,
\ee
where, in passing to the second equality, we have used 
(\ref{constraints}) and (\ref{energypreserved}).
Similarly, starting from $E_{2}\,  A^{\dagger}_{3}(p)\, B 
|0\rangle$, we obtain
\be
\frac{r_{1}}{r} = e^{i p} \frac{b(-p)}{b(p)} 
= e^{i p} \frac{c(p)}{c(-p)} 
= e^{i p} \frac{x^{+}(-p)}{x^{+}(p)} 
\frac{\gamma(p)}{\gamma(-p)}= -\frac{e^{i p}}{x^{+} 
s(x^{-})}\frac{\gamma(p)}{\gamma(-p)}\,,
\ee
where we have used (\ref{reflection}).
The same results are obtained using instead $F_{2}$.
We conclude that the boundary $S$-matrix is given (up to a scalar 
factor) by the diagonal matrix
\be
R(p) =  \diag( 
-\frac{e^{i p}}{x^{+} s(x^{-})}\frac{\gamma(p)}{\gamma(-p)} \,,
e^{-ip}\frac{\gamma(p)}{\gamma(-p)}\,,  
1 \,, 1 ) \,.
\label{boundarySRY2}
\ee
We have explicitly verified with Mathematica that this matrix satisfies the 
BYBE (\ref{BYBERY}). In the $q \rightarrow 1$ limit, (\ref{boundarySRY2})
reduces to the corresponding undeformed boundary $S$-matrix in \cite{AN}.

\subsection{$Z=0$ giant graviton brane}\label{subsec:Z}

Following \cite{HM}, we assume that the $Z=0$ giant graviton brane has a boundary degree 
of freedom and full $q$-deformed $su(2|2)$ symmetry. We therefore introduce 
a boundary operator with an index $B_{j}$,
\be
A^{\dagger}_{i}(p)\, B_{j} = R_{i\, j}^{\, i' j'}(p)\, A^{\dagger}_{i'}(-p)\,  
B_{j'} \,,
\label{boundarySRZ}
\ee
and we arrange the boundary $S$-matrix elements into the $16 \times 
16$ matrix $R$,
\be
R = R_{i\, j}^{\,i' j'} e_{i\, i'}\otimes e_{j\, j'}\,.
\label{boundSRZ2}
\ee
It satisfies the right BYBE (cf. Eq. (\ref{BYBERY}))
\be
S_{12}(p_{1}, p_{2})\, R_{13}(p_{1})\, S_{21}(p_{2}, -p_{1})\, 
R_{23}(p_{2}) = 
R_{23}(p_{2})\, S_{12}(p_{1}, -p_{2})\, R_{13}(p_{1})\, 
S_{21}(-p_{2}, -p_{1}) \,.
\label{BYBERZ}
\ee 

The vacuum state $B_{j} |0\rangle$ must form a fundamental
representation of the symmetry algebra.  The nontrivial actions of the
Cartan generators are therefore given by (cf.  Eq.  (\ref{CartanZF}))
\be
h_{1}\, B_{1} &=& -B_{1}\,, \qquad
h_{1}\, B_{2} = B_{2}\,, \qquad 
h_{3}\, B_{3} = -B_{3}\,,  \qquad
h_{3}\, B_{4} = B_{4}\,,   \non \\
h_{2}\, B_{1} &=& -\big(C_{B}-\frac{1}{2}\big) B_{1}\,, \qquad
h_{2}\, B_{2} = -\big(C_{B}+\frac{1}{2}\big) B_{2}\,,  \non \\
h_{2}\, B_{3} &=& -\big(C_{B}-\frac{1}{2}\big) B_{3}\,, \qquad
h_{2}\, B_{4} = -\big(C_{B}+\frac{1}{2}\big) B_{4}\,. 
\label{CartanZFZ}
\ee
The remaining such actions are trivial, $h_{j}\, B_{k} = 0$.
The nontrivial actions of the bosonic simple roots are given by
(cf. Eq. (\ref{bosonicsimpleZF}))
\be
E_{1}\, B_{1} = q^{1/2}\, B_{2}\,, \qquad
E_{3}\, B_{4} = q^{-1/2}\, B_{3}\,, \qquad
F_{1}\, B_{2} = q^{-1/2}\, B_{1}\,, \qquad
F_{3}\, B_{3} = q^{1/2}\, B_{4}\,, \qquad \,,
\label{bosonicsimpleZFZ}
\ee
and the remaining such actions are trivial, 
$E_{j}\, B_{k} = F_{j}\, B_{k} = 0$. Finally, 
the nontrivial actions of the fermionic generators
are given by (cf. Eq. (\ref{fermionicZF1}))
\be
E_{2}\, B_{2} =
a_{B}\, B_{4}\,,  \qquad
E_{2}\, B_{3} =
b_{B}\, B_{1}\,,  \qquad
F_{2}\, B_{1} =
c_{B}\, B_{3}\,,  \qquad
F_{2}\, B_{4} =
d_{B}\, B_{2}\,, 
\label{fermionicZF1Z}
\ee
and the remaining such actions are trivial.  The vacuum state indeed
forms a representation of the algebra (\ref{EF1}) provided the
parameters $a_{B}, b_{B}, c_{B}, d_{B}$ obey the constraints
\be
a_{B} d_{B} =\big[C_{B} + \frac{1}{2}\big]_{q}\,, 
\quad b_{B} c_{B} = \big[C_{B} - \frac{1}{2}\big]_{q}\,,
\label{constraintsZ}
\ee
which imply
\be
(a_{B} d_{B} - q b_{B} c_{B})(a_{B} d_{B} - q^{-1} b_{B} c_{B}) = 1 
\,,
\ee
in parallel with the bulk case.

A further important constraint on the parameters
$a_{B}, b_{B}, c_{B}, d_{B}$ comes from the
requirement that the central charges $P$ and $K$ (\ref{centralcharges}) 
commute with reflection from the boundary. Acting with $P$ on both 
sides of 
\be
A_{1}^{\dagger}(p)\, B_{1}\, |0\rangle =
R_{1\, 1}^{1\, 1}(p)\, A_{1}^{\dagger}(-p)\, 
B_{1}\, |0\rangle \,,
\label{11stateZ}
\ee
we obtain
\be
q^{C_{B}} e^{-i p} a(p) b(p) + q^{-C} e^{-i p} a_{B} b_{B} =
q^{C_{B}} e^{i p} a(-p) b(-p) + q^{-C} e^{i p} a_{B} b_{B} \,,
\ee
which implies
\be
a_{B} b_{B} = -g q^{C_{B}} \,.
\label{PconstraintZ}
\ee
Similarly, acting with $K$ on (\ref{11stateZ}), we obtain 
\be
c_{B} d_{B} = -g q^{C_{B}} \,.
\label{KconstraintZ}
\ee

The constraints (\ref{PconstraintZ}) and (\ref{KconstraintZ}) are 
satisfied if we set
\be
a_{B} &=& \sqrt{g}\, \gamma_{B}\, q^{C_{B}/2} \,, \non \\
b_{B} &=& -\sqrt{g}\, \frac{1}{\gamma_{B}}\, q^{C_{B}/2}  \,, \non \\
c_{B} &=& -i\sqrt{g}\, \gamma_{B}\, \frac{q^{(C_{B}+1)/2}}{x_{B}}\,, \non \\
d_{B} &=& -i\sqrt{g}\, \frac{1}{\gamma_{B}} q^{(C_{B}-1)/2} x_{B}\,,
\ee 
where $\gamma_{B}$ is left unspecified.

The constraints (\ref{constraintsZ}) then imply 
\be
q^{2C_{B}} = \frac{1}{q} \left[1 + i g (q - q^{-1}) x_{B}/q 
\right]^{-1} = 
q  \left[1 - i g (q - q^{-1}) q/x_{B} \right]^{-1}  \,.
\ee
These relations in turn imply the quadratic constraint 
\be
x_{B} + \frac{1}{x_{B}} = \frac{i}{g} \,,
\ee
which coincides with the result for the undeformed case \cite{HM}.

Having specified the representation of the boundary operator, we can
now proceed to determine the boundary $S$-matrix as before.  The
nonzero matrix elements are
\be
R_{a\, a}^{a\, a} &=& \mathcal{ A}\,, \qquad \qquad \ \
R_{\alpha\, \alpha}^{\alpha\, \alpha} = \mathcal{ D}\,, \non \\
R_{a\, b}^{b\, a} &=& \frac{\mathcal{ A}-\mathcal{ B}}{q+q^{-1}}\,, \qquad 
R_{a\, b}^{a\, b} = \frac{q^{-\epsilon_{a b}}\mathcal{ A}
+q^{\epsilon_{a b}}\mathcal{ B}}{q+q^{-1}} \,, \non \\
R_{\alpha\, \beta}^{\beta\, \alpha} &=& 
\frac{\mathcal{ D}-\mathcal{ E}}{q+q^{-1}}\,, \quad 
R_{\alpha\, \beta}^{\alpha\, \beta} = 
\frac{q^{-\epsilon_{\alpha \beta}}\mathcal{ D}
+q^{\epsilon_{\alpha \beta}}\mathcal{ E}}{q+q^{-1}} \,, \non \\
R_{a\, b}^{\alpha\, \beta} &=& 
-q^{(\epsilon_{a b}+\epsilon_{\alpha \beta})/2}
\epsilon_{a b}\epsilon^{\alpha \beta}\, \frac{\mathcal{ C}}{q+q^{-1}} \,, \quad
R_{\alpha\, \beta}^{a\, b} = 
q^{(\epsilon_{\alpha \beta}+\epsilon_{a b})/2}
\epsilon^{a b}\epsilon_{\alpha \beta}\, \frac{\mathcal{ F}}{q+q^{-1}} \,, \non \\
R_{a\, \alpha}^{a\, \alpha} &=& \mathcal{ K}\,, \quad 
R_{a\, \alpha}^{\alpha\, a} = \mathcal{ L} \,, \quad 
R_{\alpha\, a}^{a\, \alpha} = \mathcal{ G}\,, \quad 
R_{\alpha\, a}^{\alpha\, a} = \mathcal{ H} \,,  
\label{boundSZ1}
\ee
where $a\,, b \in \{1\,, 2\}$ with $a \ne b$; and 
$\alpha\,, \beta \in \{3\,, 4\}$ with $\alpha \ne \beta$; and 
\be
\mathcal{ A} &=& \frac{\gamma(p)}{\gamma(-p)}
\frac{q x^{+}+x_{B}}{x^{+} (q - x_{B} s(x^{-}))}\,, \non \\
\mathcal{ B} &=& \frac{\gamma(p)}{\gamma(-p)}\frac{x^{-}x^{+}(x_{B}+q x^{+})
+(1+q^{-2})q^{-2C}((x^{+})^{2}-(q x^{-})^{2})(q x_{B}- x^{+})}
{x^{-} (x^{+})^{2} (q - x_{B} s(x^{-}))}
\,, \non \\
\mathcal{ C} &=& \frac{q^{-5 C/2}(1+q^{-2}) \gamma_{B}\gamma(p)
\left[ q x_{B} x^{+}(1+x^{-}s(x^{+}))
+(q x^{-})^{2}-(x^{+})^{2}\right]}{x^{-} x^{+} (q - x_{B} s(x^{-}))}
\,, \non \\
\mathcal{ D} &=& 1 \,, \non \\
\mathcal{ E} &=& \frac{1}{x^{-} x^{+} s(x^{+}) (q - x_{B} s(x^{-}))}\Big\{ 
s(x^{-})\left[ q^{3}(x^{-})^{2}(1+q x_{B} s(x^{-})) +
(1+q^{2}) x_{B} x^{+} \right] \non \\
& & -(1+q^{-2}) q^{-2C} s(x^{+})\left[ q x_{B} x^{+}(1+x^{-}s(x^{+}))
+(q x^{-})^{2}-(x^{+})^{2}\right]\Big\}
\,, \non \\
\mathcal{ F} &=& \frac{1}{\gamma_{B}\gamma(-p)}
\frac{q^{-3 C/2}(1+q^{-2})}{x^{+} (x^{-})^{2} (q - x_{B} 
s(x^{-}))}\Big\{ x^{-}\left[ q x_{B} x^{+}(1+x^{-}s(x^{+}))
+(q x^{-})^{2}-(x^{+})^{2}\right]\non \\
& & -q^{2C} x_{B}\left[ q^{2} (x^{-})^{3} s(x^{+}) + (x^{+})^{2} 
\right] \Big\}
\,, \non \\
\mathcal{ G} &=& \frac{\gamma_{B}}{\gamma(-p)}
\frac{q^{(C-1)/2}((x^{+})^{2}-q^{2}(x^{-})^{2})}
{x^{+} x^{-}(q - x_{B} s(x^{-}))}
\,, \non \\
\mathcal{ H} &=& \frac{q x^{+}-x_{B} x^{-} s(x^{+})}{q x^{-} (q - x_{B} s(x^{-}))}
\,, \non \\
\mathcal{ K} &=& \frac{\gamma(p)}{\gamma(-p)}
\frac{q^{3} (x^{-})^{2} + x_{B} x^{+}}{q x^{+} x^{-} (q - x_{B} s(x^{-}))}
\,, \non \\
\mathcal{ L} &=& \frac{\gamma(p)}{\gamma_{B}}
\frac{q^{-(C+1)/2} x_{B} (1+ x^{-} s(x^{+}))}{x^{-} (q - x_{B} s(x^{-}))}
\,. \label{boundSZ2}
\ee 
We have again set the overall scalar factor equal to one.
We have explicitly verified with Mathematica that the BYBE (\ref{BYBERZ}) is
satisfied. The singularity, which in the undeformed case is at $x^{-} 
= x_{B}$, is now given by $s(x^{-}) = q/x_{B}$.

\section{Discussion}\label{sec:conclude}

We constructed the $q$-deformation of the ZF formalism developed in
\cite{AFZ, AN}, which is convenient for performing explicit
computations.  We used this formalism to reobtain the bulk $S$-matrix
of Beisert and Koroteev (\ref{bulkS3}), (\ref{bulkS4}).  We determined
how $x^{\pm}$ transforms under the reflection $p \mapsto -p$ in the
$q$-deformed theory (\ref{reflection}), and we found $q$-deformations
of the $Y=0$ and $Z=0$ giant graviton brane boundary $S$-matrices of
Hofman and Maldacena, namely, (\ref{boundarySRY2}) and
(\ref{boundSZ1}), (\ref{boundSZ2}), respectively.

It would be interesting to find additional boundary $S$-matrices,
depending perhaps on one or more boundary parameters, by looking for
linear combinations of generators which are preserved by the boundary.
Indeed, the ZF formalism is well-suited for addressing that problem.
As already mentioned in the Introduction, another interesting problem
is to construct and solve open deformed Hubbard models based on the
new boundary $S$-matrices.  Finally, pursuing the speculation in
\cite{BK} regarding a possible ``AdS${}_{q}$/CFT${}_{q}$'' duality, we
simply note that our $q$-deformed boundary $S$-matrix could then
describe the scattering of excitations of an open string attached to a
quantum-deformed giant graviton in $S^{5}_{q}$.

\section*{Acknowledgments}
One of us (RN) is grateful to C. Ahn for his collaboration on the 
related earlier project \cite{AN}.
This work was supported in part by the
National Science Foundation under Grants PHY-0244261 and PHY-0554821.

\end{document}